\documentclass[prl,twocolumn,tightenlines,preprintnumbers,floatfix,nofootinbib]{revtex4}
\usepackage{graphicx,amsfonts,amssymb,amsmath}

\renewcommand{\thefootnote}{\fnsymbol{footnote}}

\newcommand{\nc}{\newcommand}
\nc{\grad}{\nabla}  
\nc{\tr}{\mathop{\rm Tr}}
\nc{\half}{{1\over 2}}
\nc{\third}{{1\over 3}}
\nc{\be}{\begin{equation}}
\nc{\ee}{\end{equation}}
\nc{\bea}{\begin{eqnarray}}
\nc{\eea}{\end{eqnarray}}
\def\ba{\begin{eqnarray}}
\def\ea{\end{eqnarray}}

\nc{\dint}[2]{\int\limits_{#1}^{#2}}
\nc{\D}{\displaystyle}
\nc{\PDT}[1]{\frac{\partial #1}{\partial t}}
\nc{\tw}{\tilde{w}}
\nc{\tg}{\tilde{g}}
\nc{\newcaption}[1]{\centerline{\parbox{5.6in}{\caption{#1}}}}
\def\href#1#2{#2} 

\def\beq{\begin{eqnarray}}   
\def\eeq{\end{eqnarray}}

\def\lsim{\mathrel{\rlap{\lower3pt\hbox{\hskip0pt$\sim$}}
    \raise1pt\hbox{$<$}}}         %less than or approx. symbol
\def\gsim{\mathrel{\rlap{\lower4pt\hbox{\hskip1pt$\sim$}}
    \raise1pt\hbox{$>$}}}         %greater than or approx. symbol

\def\Id{\hbox{1\kern-.23em{\rm l}}}

\def\lsim{\mathrel{\rlap{\lower3pt\hbox{\hskip0pt$\sim$}}
    \raise1pt\hbox{$<$}}}         
\def\gsim{\mathrel{\rlap{\lower4pt\hbox{\hskip1pt$\sim$}}
    \raise1pt\hbox{$>$}}}         
\nc{\al}{\alpha}
\nc{\ga}{\gamma}
\nc{\de}{\delta}
\nc{\ep}{\epsilon}
\nc{\ze}{\zeta}
\nc{\et}{\eta}

\nc{\Th}{\Theta}
\nc{\ka}{\kappa}
\nc{\la}{\lambda}
\nc{\rh}{\rho}
\nc{\si}{\sigma}
\nc{\ta}{\tau}
\nc{\up}{\upsilon}
\nc{\ph}{\phi}
\nc{\ch}{\chi}
\nc{\ps}{\psi}
\nc{\om}{\omega}
\nc{\Ga}{\Gamma}
\nc{\De}{\Delta}
\nc{\La}{\Lambda}
\nc{\Si}{\Sigma}
\nc{\Up}{\Upsilon}
\nc{\Ph}{\Phi}
\nc{\Ps}{\Psi}
\nc{\Om}{\Omega}
\nc{\ptl}{\partial}
\nc{\del}{\nabla}
\nc{\ov}{\overline}
\nc{\gsl}{\!\not}
\nc{\bi}[1]{\bibitem{#1}}
\nc{\fr}[2]{\frac{#1}{#2}}
\nc{\dsl}{\partial\!\!\!\!\!\!\not\,\,}
\nc{\gm}{\mbox{$\gamma_{\mu}$}}
\nc{\gn}{\mbox{$\gamma_{\nu}$}}
\nc{\Le}{\mbox{$\fr{1+\gamma_5}{2}$}}
\nc{\Ri}{\mbox{$\fr{1-\gamma_5}{2}$}}
\nc{\GD}{\mbox{$\tilde{G}$}}
\nc{\gf}{\mbox{$\gamma_{5}$}}
\nc{\Ima}{\mbox{Im}}
\nc{\Rea}{\mbox{Re}}

\nc{\av}{\langle \ph\rangle}
\nc{\ntwo}{${\cal N}\!\!=\!2\;$}
\nc{\none}{${\cal N}\!\!=\!1\;$}
\nc{\nfour}{${\cal N}\!\!=\!4\;$}

\def \bi{\bibitem}
\nc{\rf}[1]{(\ref{#1})}
\def \del{\partial}

\renewcommand{\thefootnote}{\arabic{footnote}}
\begin{document}

\preprint{$\;\;\;$UVIC-TH-06-10}

\setcounter{page}{1}

%%
%% The title page
%% 
%\begin{titlepage}
\renewcommand{\thefootnote}{\fnsymbol{footnote}}

\setcounter{page}{1}

%\vspace*{0.1in}

\hspace{-4cm}\title{$\;\;\;\;\;\;\;\;\;\;\;\;\;$
Electric Dipole Moments as Probes of $CPT$ Invariance $\;\;\;\;\;\;\;\;\;\;\;\;\;$}

\author{Pavel A. Bolokhov$^{\,(a,b)}$, Maxim Pospelov$^{\,(a,c)}$ 
 and Michael Romalis$^{\,(d)}$}

\affiliation{$^{\,(a)}${\it Department of Physics and Astronomy, University of Victoria, 
     Victoria, BC, V8P 1A1 Canada}\\
$^{\,(b)}${\it Theoretical Physics Department, St.Petersburg State University, Ulyanovskaya 1,
        Peterhof, St.Petersburg, 198504, Russia}\\
$^{\,(c)}${\it Perimeter Institute for Theoretical Physics, Waterloo,
Ontario N2J 2W9, Canada}\\
$^{\, (d)}${\it Department of Physics, Princeton University, Princeton, New Jersey, 08550, USA}}

%\centerline{\large\bf Abstract}
\begin{abstract}
Electric dipole moments (EDMs) of elementary particles and atoms probe violations of 
$T$ and $P$ symmetries and consequently of $CP$ if $CPT$ is an exact symmetry. 
We point out that EDMs can also serve as sensitive probes of $CPT$-odd, $CP$-even
interactions, that are not constrained by any other existing experiments.  
Analyzing models with spontaneously broken Lorentz invariance, 
we calculate EDMs in terms of the leading $CPT$-odd operators to show that
experimental sensitivity probes the scale of $CPT$ breaking as high as $10^{12}$ GeV. 
\end{abstract}

\maketitle

%\tighten

%%%%%%%%%%%%%%%%%%%%%%%%%%%%%%%%%%%%%%%%%%%%%%
%%%%
%%%               Introduction
%%%%
%%%%%%%%%%%%%%%%%%%%%%%%%%%%%%%%%%%%%%%%%%%%%%

%\section{Introduction}

Tests of fundamental symmetries play an important 
role in discerning the properties of nature at 
ultra-short distance scales. Initially suggested as 
an accurate test of parity conservation in strong interactions 
\cite{PR}, the electric dipole moments (EDMs) of neutrons and heavy atoms 
provide an important test of $P$ and $T$ symmetries \cite{exp,KL,PRann}. 
A non-relativistic Hamiltonian for a neutral particle of spin 
$S$ can be written as the combination of two terms,
\ba
 H = - \mu {\bf B} \cdot \fr{\bf{S}}{S} - d {\bf E} \cdot \fr{\bf{S}}{S}\ .
\label{starting}
\ea
Under the reflection of spatial coordinates, $P({\bf B} \cdot \bf{S})= {\bf B} \cdot \bf{S}$, 
whereas $P({\bf E} \cdot \bf{S})= -{\bf E} \cdot \bf{S}$. Under time reflection, 
$T({\bf B} \cdot \bf{S})= {\bf B} \cdot \bf{S}$ and 
$T({\bf E} \cdot \bf{S})= -{\bf E} \cdot \bf{S}$. The presence of a non-zero $d$
would therefore signify the existence of both parity and time-reversal
violation. In a world with perfect $CPT$ symmetry, a search for $d$ would also 
be a direct test of $CP$ symmetry. 
%In particular, it means that the 
%product $\mu d$ will be of the same sign for particle and antiparticle. 
An assumption of $CPT$ is well-justified 
in the field theory framework, as it rests on locality, spin-statistics and
Lorentz invariance. Nevertheless, independent tests of $CPT$ are warranted, and a
number of searches in the $K$ and $B$ meson systems \cite{K_CPT},
as well as with electrons, muons and antiprotons have been pursued over the years. 
In this paper, we show that EDMs can serve as a sensitive 
probe of $CPT$ violation, independent from other available tests.
More specifically, we  relax the assumption of Lorentz invariance 
thus enabling the breaking of $CPT$ and study the 
EDMs induced by $CPT$-odd but $CP$-even interactions. 
%This makes the application of current EDM constraints 
%and interpretation of future EDM results more generic, providing complimentary 
%constraints on parameters of $CPT$ violation. 
%n addition, we would like to show how the 
%$CPT$-odd and $CP$-odd effects can be distinguished, should the next generation
%of EDM experiments find a non-zero result.

Suppose that the breaking of $CPT$ symmetry comes from 
some unknown, presumably short-distance scale physics and 
manifests itself in the interaction of Standard Model fields with
external backgrounds that transform as vectors and tensors 
under the Lorentz group \cite{Kost,CG}. The simplest possibility
is to have a time-like condensation of a vector $n_\mu = (1,0,0,0)$
that introduces a preferred frame. For simplicity we assume that 
$n_\mu $ coincides with the laboratory frame, but the results can be easily generalized for
a generic frame. In the presence of such a vector, 
the EDM part of Hamiltonian (\ref{starting}) for the spin 1/2 particle can be rewritten 
as 
\be
{\cal L}_{\rm EDM} = \fr{-i}{2}d_{\rm CP}\ov \psi \sigma^{\mu\nu}F_{\mu\nu} \psi 
+ d_{\rm CPT}\ov{\psi} \gamma_\mu\gamma_5 \psi F_{\mu\nu}n^{\nu}
\label{start}
\ee
where $d_{\rm CP}+d_{\rm CPT}=d$. 
Thus, quite generically, 
the nil result for the neutron EDM searches provides a constraint on the combination 
$d_{\rm CP}+d_{\rm CPT}$.
Introducing an axial four-vector of spin $ \mathfrak{a}^\mu $ and four-velocity $u^\mu$, 
we generalize (\ref{start}) for a particle of arbitrary spin:
%by introducing the spin 4-vector $ a^\mu $ (which corresponds to the axial current in the
%case of the spin-1/2 particle)
\begin{equation}
\label{arb_spin}
{\cal L}_{\rm EDM} =F_{\mu\nu} \mathfrak{a}^{\nu} (d_{\rm CP}\,u^\mu
	~+~
	d_{\rm CPT}\, n^\mu ).
\end{equation}
Allowing for more complicated backgrounds, we notice that the $CPT$-odd EDM-type correlation
may also result from interaction with irreducible tensor $D^{\mu\nu\rho}$, symmetric in 
$\nu\rho$: $F_{\mu\nu} \mathfrak{a}^{\rho}D^{\mu\nu\rho}$.
In the remainder of this paper, we analyze the structure of the 
$CPT$-odd and $CP$-even effective Lagrangian, deduce its consequences 
for the EDMs of neutrons and heavy atoms, and explore the possibility of 
distinguishing 
$d_{\rm CP}$ and $d_{\rm CPT}$ in experiment, should the non-zero EDMs be found. 

%\section{$CPT$-odd, $CP$-even interactions}
{\em $CPT$-odd, $CP$-even operators.} In the framework where $CPT$ violation 
is mediated by Lorentz violation, the
$CPT$-odd interaction terms appear at odd dimensions \cite{Kost}. 
All $CPT$-odd dimension three operators can be easily listed \cite{Kost}, 
\be
{\cal L}_3 = -\sum \bar \psi( a^\mu \gamma_\mu + b^\mu \gamma_\mu \gamma_5)\psi,
\label{dim3}
\ee
with $a_\mu$ and $b_\mu$ being Lorentz/CPT violating couplings with possible flavor dependence. 
%In this paper, for simplicity, we would 
%assume the presence of only one vector background, and take $a_\mu = a n_\mu$, $b_\mu = 
%b n_\mu$. 
Only certain types of $CPT$-violating dimension five operators
were classified in the literature \cite{MP}, and here we complement this list by 
including operators linear in the gauge field strength:
%can be easily listed:
\ba
\mathcal{L}_{ 5} =  -\sum [\nonumber
	c^\mu \overline{\psi} \gamma^\lambda F_{\lambda\mu} \psi
	~+~
	d^\mu \overline{\psi} \gamma^\lambda \gamma^5 F_{\lambda\mu} \psi
	\\~+~
{f}^\mu \overline{\psi} \gamma^\lambda \gamma^5
	\widetilde{F}_{\lambda\mu} \psi
	~+~
{g}^\mu  \overline{\psi} \gamma^\lambda 
	\widetilde{F}_{\lambda\mu}  \psi].
	\label{dim5}
\ea
The sum spans different fermions of the SM and different gauge groups, 
with $F_{\mu\nu}$ standing for the corresponding field strength. 
We note that, as it is usual in such theories, the LV theory described by interactions \eqref{dim3} 
and \eqref{dim5} is considered a ``safe'' effective low-energy description of the unknown UV physics.
The UV theory is assumed to be Lorentz invariant, and therefore the effective theory is not
expected to suffer from any conceptual issues related to broken Lorentz invariance, such as
{\it e.g.} violation of microcausality.
Assuming that the vector backgrounds are time-like and invariant under $C$, $P$ and $T$ reflections, 
we classify the properties of operators (\ref{dim3}) and (\ref{dim5}) under these 
discrete symmetries in Table 1. There is only one operator 
that is odd under parity and time reversal, and thus our further analysis concentrates
only on $d^\mu$. 

\begin{table}[tb]
\label{CPT_table}
\begin{equation*}
\begin{array}{cc||ccc}
\hline
 
${\rm Coefficient}$ & ${\rm Operator}$   &  C  &  P  &  T    \\
     
\hline
a^0 &\overline{\psi}\gamma_0 \psi & - & + & +\\

b^0 &\overline{\psi}\gamma_0 \gamma_5\psi & + & -& +\\

c^0 &F_{\lambda 0} \overline{\psi} \gamma^\lambda  \psi &
	~~+~~  &  ~~+~~  &  ~~-~~    \\

d^0 &F_{\lambda 0}  \overline{\psi} \gamma^\lambda \gamma^5 \psi &
	-  &  -  &  -    \\
	
f^0 &\widetilde{F}_{\lambda 0} \overline{\psi} \gamma^\lambda 
	 \gamma^5 \psi &
	-  &  +  &  +    \\
	
g^0 &\widetilde{F}_{\lambda 0} \overline{\psi} \gamma^\lambda 
	 \psi &
	+  &  -  &  +    \\

\hline
\end{array}	
\end{equation*}
\caption{$C$, $P$, $T$ properties of dimension three and five Lorentz violating $CPT$-odd
 operators. Only one operator
proportional to $d^0$ is both $P$ and $T$ odd and capable of inducing EDMs. }
\end{table}

It is convenient to classify these operators at  
the scale of 1 GeV, where only light quark fields, gluons,  photons,
electrons and muons are the remaining degrees of freedom, while weak bosons and 
heavy quarks are already decoupled. 
Taking a quark field $\psi_q$ with the electric charge $Q_q$, 
and using the full equation of motion in the electromagnetic 
and strong backgrounds, 
\be
iD_\mu\gamma^\mu \psi_q \equiv  (i\partial_\mu - g_st^a A^a_\mu - eQ_q A_\mu)\gamma^\mu
\psi_q = m_q \psi_q,
\ee 
we deduce an identity that relates gluon and 
photon-containing operators for quarks: 
\begin{align}
&\bar \psi_q(eQ_q F_{\mu\nu} + g_st^aG_{\mu\nu}^a)\gamma^\nu \gamma_5\psi_q = 
-i\bar \psi_q[D_\mu,D_\nu\gamma^\nu\gamma_5] \psi_q \nonumber\\
&\qquad =2m_i \bar \psi_q D_\mu \gamma_5 \psi_q = m_q \bar \psi_q
[ D_\nu \gamma^\nu, \gamma_\mu\gamma_5 ] \psi_q=0.
\label{relation}
\end{align}
Here $[,]$ is the commutator.  Eq. (\ref{relation}) effectively 
reduces the number of independent quark operators, and we choose to  eliminate
$\ov{\psi}{}_qg_st^aG_{\mu\nu}^a\gamma^\nu \gamma_5\psi_q$ by expressing it via 
$\ov{\psi}{}_qeQ_q F_{\mu\nu} \gamma^\nu \gamma_5\psi_q$.
 Remarkably, there is no $CPT$-odd, $CP$-even operators for Dirac particles that have only 
electromagnetic interactions, such as muons and electrons, because in this case  Eq. (\ref{relation})
degenerates to an identity $\ov{\psi}{}_eF_{\mu\nu} \gamma^\nu \gamma_5\psi_e = 0 $. It turns out that the 
vanishing of this effective operator is well known in the standard $CP$-odd EDM computations. 
The correction to the electron Hamiltonian created by operator 
$\ov{\psi}{}_eF_{0\nu} \gamma^\nu \gamma_5\psi_e$ is proportional to the product of electric field 
and relativistic spin operator $\bf{\Sigma}$, $\bf{ E \Sigma}$. This product can be 
represented as a result of the commutator of another operator with 
the full  Dirac Hamiltonian, $\bf{ E \Sigma}$$ = (1/e)$$ [\bf{\Sigma\nabla}$$, H]$. Therefore, 
the expectation value of $\bf{ E \Sigma}$ over any eigenstate of $H$ is zero \cite{KL,FG},
which is another way of stating that $\ov{\psi}{}_eF_{0\nu} \gamma^\nu \gamma_5\psi_e$ vanishes on shell.

%The vanishing of
%$\bar \psi_e F_{\mu\nu}\gamma^\nu \gamma_5\psi_e$ on the equation of motion 
%means that lepton's trajectory is unaffected by such operator 
%provided that the electron (muon) dynamics is completely determined by 
%electromagnetic interaction. 
%An inclusion of weak interactions would result 
%in a non-zero effect, but will be additionally suppressed by the Fermi constant $G_F$ 
%and thus can be absorbed into higher-dimensional operators. 
Taking these identities into account, we write down 
the effective $T$, $P$, $CPT$-odd Lagrangian at 1 GeV scale in a remarkably simple form,
 that contains only three terms:
\be
{\cal L}_{\rm CPT} = \sum_{i=u,d,s} d^\mu_i \bar q_i \gamma^\lambda \gamma^5 
F_{\lambda\mu} q_i.
\label{CPT}
\ee
This is a rather compact form compared to a usual $CP$-odd effective Lagrangian 
where a few dozens of terms have to be taken into account \cite{PRann}. 

An important difference between $CP$-odd and $CPT$-odd EDMs comes from the 
$SU(2)\times U(1)$ properties of Eq. (\ref{CPT}). 
%allows to determine the decoupling properties of these interactions with respect to 
%the energy scale of $CPT$ violation $\Lambda_{\rm CPT}$. 
$CP$-odd effects require helicity flip and thus correspond to dimension 6 operators
above the electroweak scale, decoupling as $1/\Lambda^2_{\rm CP}$ as the scale of 
$CP$ violation $\Lambda_{\rm CP}$ gets larger. 
One can easily see that $CPT$-odd terms (\ref{CPT}) 
%and (\ref{CPT1}) 
correspond to genuine dimension 
5 operators such as $\bar q_{R(L)} \gamma^\lambda \gamma^5 F_{\lambda\mu} q_{R(L)}$
and $\bar q_L \gamma^\lambda \gamma^5 \tau^a F^a_{\lambda\mu} q_L$ and do not
require chirality flip.  
%It is important 
%to recall that the conventional $CP$-odd EDMs $d_{\rm CP}$ 
%mix left- and right-handed fermions and require the 
%Higgs vev insertion, so that $d_{\rm CP} \propto m_q\Lambda_{\rm CP}^{-2}$,
%while the 
Consequently, $CPT$-odd physics decouples only linearly, $d_{\rm CPT} \propto 
\Lambda_{\rm CPT}^{-1}$. Combination of present day limit on neutron EDM with the
linear decoupling property furnishes the sensitivity to the scales of $CPT$ violation as 
large as
\be
\Lambda_{\rm CPT} \sim (10^{11}-10^{12})~{\rm GeV}.
\label{range}
\ee
Future generation experiments could 
potentially probe $CPT$-violating physics all the way to the Planck 
scale, being limited only by the prediction of the Kobayashi-Maskawa (KM) model for the 
neutron EDM at the level of $10^{-31}-10^{-33}e$cm.

%\section{Signatures of $CPT$-odd EDM's}

{\em Signatures of $CPT$-odd EDMs. }
There are three main groups of observable EDMs, which include EDMs of neutrons, 
diamagnetic atoms (Hg, Xe, etc.) and paramagnetic  atoms (Tl, Cs, etc.). 
%The prediction for the neutron EDM $d_n$ in terms of 
%$d_{u,d,s}^\mu$ differs
%slightly from $d_n$ induced by the $CP$-odd EDMs of quarks. 
A rather simple structure of the $CPT$-odd effective Lagrangian 
helps to determine the dependence of these
observables on different $d_i^\mu$ in  (\ref{CPT}).

The QCD calculations of conventional $CP$-odd EDMs \cite{PRann}
are very close 
to a constituent quark model prediction, $d_n \simeq \fr{4}{3} d_d- \fr{1}{3} d_u$,
with the contribution of the $s$-quark being zero. In the $CPT$-odd case, we use 
matrix elements of the axial-vector charges of light quarks inside a nucleon, 
which can be obtained from the nucleon spin structure functions \cite{EK}. 
This way, to $\sim$20\% accuracy, we get
\be
d_n \simeq 0.8 d^0_d - 0.4 d^0_u - 0.1 d^0_s.
\label{dn}
\ee
Using $|d_n| < 3\times 10^{-26} e$cm \cite{exp}
and barring significant cancellation between the constituents, 
we conclude that $CPT$-odd EDMs of light quarks are limited 
at $O(10^{-25}e$cm$)$.

The measurements of EDMs of diamagnetic atoms are usually 
quite competitive with $d_n$ due to 
color EDM contributions to the $CP$-odd pion-nucleon 
coupling constant $\bar g_{\pi NN}$ \cite{KL,PRann}. 
As we already noted, interactions (\ref{CPT}) preserve quark chirality, and 
involve a photon field, thus leading to a strong suppression of $\bar g_{\pi NN}(d^\mu_q)$, 
which makes the $T$-odd pion exchange ineffective. 
Consequently, the EDM of the diamagnetic atoms are induced by
the EDMs of the valence nucleons.
For the most important case of mercury EDM \cite{DS}, we have
\ba
d_{\rm Hg} \simeq - 5\times 10^{-4}(d_n + 0.1 d_p)\nonumber\\\simeq
-5\times 10^{-4}(0.74 d_d^0 -0.32 d_u^0 - 0.11 d_s^0),
\label{dHg}
\ea
and an approximate relation $d_{\rm Hg}/d_n \sim - 5\times 10^{-4}$ could be 
interpreted as a signal consistent with $CPT$ violation should the nonzero 
$d_{\rm Hg}$ and $d_n$ be found. Due to absence of $CPT$-odd electron 
EDM operator, EDMs of paramagnetic atoms are predicted to be extremely suppressed.

%A suggested experiment for deuteron EDM \cite{Deut}, 
%would provide a sensitivity to an 
%orthogonal combination of $d_q$,
%$d_D = 0.4 d_d + 0.4 d_u - 0.2 d_s$. 

An unambiguous separation of $CP$-odd and $CPT$-odd EDM terms in (\ref{arb_spin}) 
may come from measuring the difference of their relativistic effects. 
%Suppose that a charged particle 
%with no anomalous magnetic moment, and with EDMs given by (\ref{arb_spin}) is moving in a 
%uniform magnetic field. 
The $CP$-odd EDM interacts with the magnetic field 
and leads to the precession of the spin relative to $[{\bf B}\times{\bf v}]$, while the 
$CPT$-odd component does not contribute to the precession for a particle on a circular orbit. 
Thus, the experimental proposal of measuring deuteron EDM in the storage ring \cite{Deut} 
would in principle have capabilities of separating the two effects, as perpendicular ${\bf B}$  and ${\bf E}$ 
would be employed in the experimental set-up. In practice, the signal of  
spin precession due to the $ CPT $-odd EDM is not exactly zero but 
suppressed by the deuteron anomaly, $ |a_D| = 0.143$, 
because of the $|{\bf E}|=|a_D{\bf B}|$ choice \cite{Deut}. The suppression of the deuteron $d_{\rm CPT}$ signal 
measured in the storage ring relative to $d_n$ is opposite to the case of $ d_{\rm CP} $ where an
enhancement of $d_D/d_n \sim 5 $ is expected \cite{LOPR} due to the $CP$-odd pion exchange.

%However, the EDM searches \cite{Deut} will measure the resonant precession of the 
%spin of deuterons in a storage ring, 
%However, deuterons in a storage ring 
%%are sensitive to the sum of both
%%contributions

{\it Naturalness.}
Since there are many other observables sensitive to Lorentz/CPT violation given by dimension 
three operators (\ref{dim3}), 
it is important to investigate whether operators of dimension five (\ref{dim5}) may 
influence these observables through quantum loops. It is easy to see, for example, 
that the last dimension five operator in (\ref{dim5}), $g^\mu \widetilde{F}_{\mu\nu} \overline{\psi}
\gamma_\nu\psi$, produces a quadratically divergent 
result for dimension three term, $b^\mu \ov{\psi} \gamma_\mu\gamma_5 \psi$, already at one loop. 
Even with a modest choice of 
the cutoff, the contribution to $b_\mu$ will 
significantly exceed present experimental bounds of order $10^{-31}$ GeV, modulo an extreme fine-tuning. 
%The most important are the dimension 3 operators, 
%and $b^\mu$-proportional terms in particular, whose spatial components can be 
%probed with $O(10^{-31})$ GeV sensitivity in the clock comparison
%experiments. 
It turns out that EDM operators $d^\mu$ are protected 
against transmutation to $a_\mu$ and $b^\mu$ to a high loop order because of their
difference in $CP$. Thus, only loops with intrinsic 
$CP$ violation can convert $d^\mu$ into $a^\mu$ or $b^\mu$. In the SM this is rather difficult 
to achieve, as the violation of $CP$ symmetry in the flavour-conserving channel 
happens minimum at three loops,
 and is further suppressed by the Kobayashi-Maskawa mixing angles and 
quark Yukawa couplings.
%Therefore, one would typically expect that for a $d$-quark
%\be
%a^\mu_d, ~b^\mu_d \sim d^\mu_d m_c^2 \times 10^{-5} \times 10^{-10},
%\label{bdmu}
%\ee
%where the two numerical factors account for the smallness of the mixing angles and 
%the loop suppression. Assuming that $d^\mu$ is time-like and fixed in the 
%cosmic or the Solar system frame, one further expects a suppression of 
%spatial components by $v_{\rm earth}/c \sim 10^{-4}-10^{-3}$. Thus our prediction for 
%anisotropic effects created by the combination of $CPT$-odd EDMs and $CP$-odd 
%radiative corrections are at the level of 
A crude estimate of dimension three operators resulting from multi-loop $CP$-violating
corrections gives an admittedly imprecise prediction for a light quark,
\be
a^\mu,b^\mu \sim d^\mu (10^{-20}-10^{-18})\times {\rm GeV}^2.
\label{bdi}
\ee
This provides sensitivity to $d_\mu$ up to $10^{-12}~{\rm GeV}^{-1}$,
which is essentially the same sensitivity as (\ref{range}). Therefore a detectable signal 
from the $CPT$-odd EDMs induced by a vector background would likely come accompanied 
by $b^\mu$, which could be searched for via {\em e.g.}
sidereal modulation of spin precession frequencies \cite{Kost}. 
A difference of down and strange $a_\mu$ terms 
can be searched for with the neutral $K$ mesons producing
a typical bound on $|a^0_s-a^0_d|$ of order $\sim 10^{-19}-10^{-20}$ GeV. 
Through the loop effects, this amounts to sensitivity to $d^0_q$ terms 
on the order of $10^{-5}$ GeV$^{-1}$, 
which is significantly less sensitive than (\ref{range}). 

{\em Tensor backgrounds}. What if the nature of $CPT$-violation is so intricate as to give rise to
an external rank-three tensorial background $D^{\mu\nu\rho}$?
 In this case the $T$, $P$ and $CPT$ odd interaction  
$F_{\mu\nu} \mathfrak{a}^{\rho}D^{\mu\nu\rho}$
induces the EDM-like signatures via an anisotropic effective Hamiltonian for the spin:
\be
H = - \mu {\bf B} \cdot \fr{\bf{S}}{S} - {\cal D}^{ij} E_i \cdot \fr{S_j}{S}.
\label{Htensor}
\ee
Here ${\cal D}^{ij}$ is the traceless symmetric tensor with spatial components,
${\cal D}^{ik} ~=~ {\cal D}^{i[0k]} ~+~ {\cal D}^{k[0i]}$. The tensor interaction in (\ref{Htensor})
creates a correction to the spin precession frequency proportional to 
$E_iB_k{\cal D}^{ik}$ which changes sign under the reversal of the electric field. 
The effect averages to zero if the orientation of 
parallel ${\bf E}$ and ${\bf B}$ fields is randomly changing relative to the 
external tensor  ${\cal D}^{ik}$ due to its tracelessness. However, in EDM experiments
such averaging is not done. Therefore, $E_iB_k{\cal D}^{ik}$ gives an EDM 
signature, which in addition changes during the day because of the change of the 
orientation of a laboratory relative to ${\cal D}^{ik}$ if, of course, the frame that breaks 
Lorentz invariance is not related to the Earth itself. Generically, 
one expects $12$ and $24$ hour modulations of the EDM signal due to 
the $CPT$-odd tensor background. 
The structure of operators leading to (\ref{Htensor})
 is more complex than in the vector case. 
In particular, the electron operator, $\bar e  F_{\mu\nu}\gamma_\rho\gamma_5   e D^{\mu\nu\rho}$ 
does not vanish, and leads to the EDMs of a paramagnetic atom, albeit with the matrix element suppressed
by a factor of $\sim 10$ relative to the $CP$-odd case.  As in the vector case, 
the EDMs of diamagnetic atoms are induced by the 
EDMs of valence nucleons. Finally, tensor backgrounds are protected against transmutation 
to lower dimensional operators. 

{\em In conclusion}, we point out that EDMs put stringent limits on a new type of 
$CPT$-odd $CP$-even interactions that is not constrained by other tests
of Lorentz invariance and $CPT$. The scale of $CPT$-breaking probed by 
current versions of EDM experiments is as high as $10^{12}$ GeV. 
The unambiguous separation of $CPT$-odd and $CP$-odd 
effects would require EDM experiments with antiparticles, which might be a formidable 
challenge. Instead, we point out the main pattern in EDM observables consistent with $CPT$
violation: nuclear and atomic EDMs will be induced by the EDMs of neutrons and protons, 
while electron EDM and $T$-odd nuclear forces are largely ineffective in the $CPT$-odd case.


\begin{thebibliography}{99}
\bi{PR} E. M. Purcell and N. F. Ramsey, Phys. Rev. {\bf 78}, 807 (1950).

\bibitem{exp}
P.~G.~Harris {\it et al.},
Phys.\ Rev.\ Lett.\  {\bf 82} (1999) 904; 
M.~V.~Romalis, W.~C.~Griffith and E.~N.~Fortson,
Phys.\ Rev.\ Lett.\  {\bf 86} (2001) 2505;
B. C. Regan {\em et al.},  Phys. Rev. Lett. {\bf 88} (2002) 
071805.

\bi{KL} I.B. Khriplovich and S.K. Lamoreaux, {\it ''CP Violation 
Without Strangeness''}, Springer, 1997;
J.~S.~M.~Ginges and V.~V.~Flambaum,
%``Violations of fundamental symmetries in atoms and tests of unification
%theories of elementary particles,''
Phys.\ Rept.\  {\bf 397}, 63 (2004).

\bi{PRann}M.~Pospelov and A.~Ritz,
  %``Electric dipole moments as probes of new physics,''
  Annals Phys.\  {\bf 318}, 119 (2005).

\bi{K_CPT} S. Eidelman et al., Phys. Lett. B592, 1 (2004); V.~A.~Kostelecky,
  %``Sensitivity of CPT tests with neutral mesons,''
  Phys.\ Rev.\ Lett.\  {\bf 80}, 1818 (1998).


\bibitem{Kost} D.~Colladay and V.A.~Kostelecky, Phys.\ Rev.\ {\bf
D55}, 6760 (1997), Phys. Rev. D {\bf 58}, 116002 (1998)% [hep-ph/9703464].
%%CITATION = HEP-PH 9703464;%%

\bibitem{CG}
S.R.~Coleman and S.L.~Glashow, Phys.\ Rev.\ {\bf D59}, 116008
(1999). % [hep-ph/9812418].


\bibitem{MP} R.~C.~Myers and M.~Pospelov,
  %``Experimental challenges for quantum gravity,''
  Phys.\ Rev.\ Lett.\  {\bf 90}, 211601 (2003);
S.~Groot Nibbelink and M.~Pospelov,
  %``Lorentz violation in supersymmetric field theories,''
  Phys.\ Rev.\ Lett.\  {\bf 94}, 081601 (2005);
  %%CITATION = HEP-PH 0404271;%%
P.~A.~Bolokhov, S.~G.~Nibbelink and M.~Pospelov,
  %``Lorentz violating supersymmetric quantum electrodynamics,''
  Phys.\ Rev.\ D {\bf 72}, 015013 (2005).
  %%CITATION = HEP-PH 0505029;%%
%P.~A.~Bolokhov, S.~G.~Nibbelink and M.~Pospelov,
%  arXiv:hep-ph/0505029.

\bibitem{FG}
  J.~S.~M.~Ginges and V.~V.~Flambaum,
   ``Violations of fundamental symmetries in atoms and tests of unification
  %theories of elementary particles,''
  Phys.\ Rept.\  {\bf 397}, 63 (2004).


\bibitem{EK} J.~R.~Ellis and M.~Karliner,
  %``Nucleon spin,''
  arXiv:hep-ph/9510402.



\bi{DS}
V.~F.~Dmitriev and R.~A.~Sen'kov,
Phys.\ Rev.\ Lett.\  {\bf 91}, 212303 (2003).


\bi{Deut}
Y.~K.~Semertzidis {\it et al.}  [EDM Collaboration],
arXiv:hep-ex/0308063.
%%CITATION = HEP-EX 0308063;%%

\bi{LOPR}O.~Lebedev, K.~A.~Olive, M.~Pospelov and A.~Ritz,
  %``Probing CP violation with the deuteron electric dipole moment,''
  Phys.\ Rev.\ D {\bf 70}, 016003 (2004).

\end{thebibliography}
\end{document}